\documentclass[prl,aps,showpacs,amsmath,amssymb,superscriptaddress,twocolumn]{revtex4-1}
\usepackage{graphicx}
\usepackage{CJK}
\usepackage{epstopdf}

\usepackage{amsmath,amssymb}
\usepackage{color}

\newcommand{\wpcm}{W/cm$^2$}

\begin{document}
\begin{CJK*}{UTF8}{gbsn}

\title{Localizing High-Lying Rydberg Wave Packets with Two-Color Laser Fields}

\author{Seyedreza Larimian}
\affiliation{Photonics Institute, Technische Universit\"at Wien, A-1040 Vienna, Austria}
\author{Ji-Wei Geng (耿基伟)}
\affiliation{State Key Laboratory for Mesoscopic Physics and Collaborative Innovation Center of Quantum Matter, School of Physics, Peking University, Beijing 100871, China}
\author{Stefan Roither}
\author{Daniil Kartashov}
\author{Li Zhang (张丽)}
\affiliation{Photonics Institute, Technische Universit\"at Wien, A-1040 Vienna, Austria}
\author{Mu-Xue Wang (王慕雪)}
\affiliation{State Key Laboratory for Mesoscopic Physics and Collaborative Innovation Center of Quantum Matter, School of Physics, Peking University, Beijing 100871, China}
\author{Qihuang Gong (龚旗煌)}
\author{Liang-You Peng (彭良友)}
\email[Electronic address: ]{liangyou.peng@pku.edu.cn}
\affiliation{State Key Laboratory for Mesoscopic Physics and Collaborative Innovation Center of Quantum Matter, School of Physics, Peking University, Beijing 100871, China}
\affiliation{Collaborative Innovation Center of Extreme Optics, Shanxi University, Taiyuan, Shanxi 030006, China}
\author{Christoph Lemell}
\author{Shuhei Yoshida}
\author{Joachim Burgd\"orfer}
\affiliation{Institute for Theoretical Physics, Technische Universit\"at Wien, A-1040 Vienna, Austria}
\author{Andrius Baltu\v{s}ka}
\author{Markus Kitzler}
\affiliation{Photonics Institute, Technische Universit\"at Wien, A-1040 Vienna, Austria}
\author{Xinhua Xie (谢新华)}
\email[Electronic address: ]{xinhua.xie@tuwien.ac.at}
\affiliation{Photonics Institute, Technische Universit\"at Wien, A-1040 Vienna, Austria}
\affiliation{Institute of Theoretical Chemistry, University of Vienna, A-1010 Vienna, Austria}
\date{\today}

\begin{abstract}
We demonstrate control over the localization of high-lying Rydberg wave packets in argon atoms with phase-locked orthogonally polarized two-color (OTC) laser fields. With a reaction microscope, we measured ionization signals of high-lying Rydberg states induced by a weak dc field and black-body radiation as a function of the relative phase between the two-color fields. We find that the dc-field ionization yields of high-lying Rydberg argon atoms oscillate with the relative two-color phase with a period of $2\pi$ while the photoionization signal by black-body radiation shows a period of $\pi$. These observations are a clear signature of the asymmetric localization of electrons recaptured into high-lying Rydberg states after conclusion of the laser pulse and are supported by a semiclassical simulation of argon-OTC laser interaction. Our findings thus open an effective pathway to control the localization of high-lying Rydberg wave packets.
\end{abstract}

\maketitle
\end{CJK*}

Highly excited Rydberg atoms and molecules, in comparison with ground-state atoms and molecules, have unique properties \cite{gallagher2005rydberg}. Such atoms and molecules can be exploited in the studies of the quantum phenomena and the transition from the quantum to the classical world on a macroscopic length scale. They play an important role in chemistry and astrophysics and are considered to be building blocks for future applications on quantum information, chemistry and astrophysics \cite{Saffman2010,*Merkt1997,*Gnedin2009}. Manipulating electrons in the ground and excited states of an atom or a molecule is of fundamental interest for physics and chemistry with a wide range of applications from high harmonic generation~\cite{Winterfeldt2008} to the control of chemical reactions~\cite{Carley2005}. Previous studies found that high-lying Rydberg states can be steered by weak half-cycle pulses~\cite{Dunning2009}.

In a strong laser pulse, valence electrons of an atom or a molecule can be detached through tunneling or barrier suppression ionization. After conclusion of the pulse, some of the released electrons may be recaptured by the ionic Coulomb field and populate highly excited Rydberg states (``frustrated field ionization'')~\cite{Krausz2009}. Recently, we have reported on the lifetime of such states measured by electron-ion coincidence spectroscopy~\cite{Larimian2016}. It has been demonstrated that electronically excited states play an important role in strong field phenomena, including ionization and molecular dissociation~\cite{Wolter14,*li14,*liu12,minns14}, electron wave packet interference, and high harmonic generation~\cite{chini2014coherent,xie12prl,xie15prl,*Deng2015,arbo14}. Many strong field phenomena in atoms and molecules are governed by electronic dynamics that are not only sensitive to the laser intensity but also to the waveform of the laser field~\cite{baltuska03:nature}. The latter can be controlled by varying the carrier-envelope phase of a few-cycle laser pulse or by the superposition of phase locked pulses with different colors~\cite{baltuska03:nature,Chan2011}.

\begin{figure}[htbp]
\centering
\includegraphics[width=0.4\textwidth,angle=0]{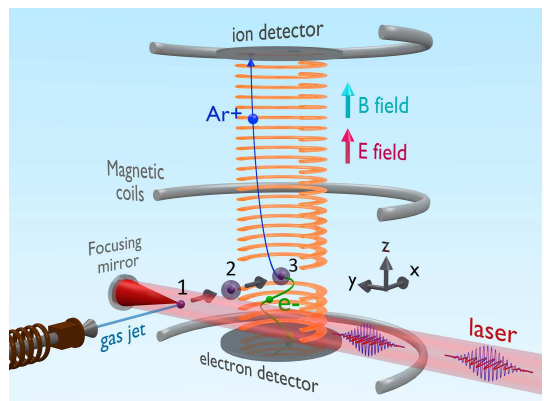}
\caption{(color online). Schematic view of the experiment. An electron from an argon atom released by the field of an OTC laser pulse may be recaptured into a high-lying Rydberg state by frustrated field ionization. The Rydberg atom, in turn, is subsequently ionized either by the weak dc field in the target region or by photoionization by black-body radiation.} \label{fig:peda}
\end{figure}
In this Letter, we report on the control of the formation of spatially localized high-lying Rydberg wave packets by waveform controlled orthogonally polarized two-color (OTC) laser fields in argon atoms. With the help of semiclassical electron-trajectory simulations we analyze the experimental observation and identify the underlying mechanism. In the present experiment we exploit the relative phase of OTC laser fields to achieve temporal and spatial shaping of the waveform of the laser field. Previously, OTC fields have been successfully proposed and applied to control electron rescattering and interference~\cite{kitzler05prl,*kitzler07pra,xie15prl,richter15,geng2015}, to image atomic wave functions based on high harmonic generation~\cite{kitzler08njp,*shafir09}, and to control electron emission and correlation in single and double ionization of atoms~\cite{zhang14prl,*zhou11,*zhang14pra}. Since controlling the wave form of an OTC laser field provides the capability of manipulating electron trajectories in time and space, one may expect to achieve control over the formation of high-lying Rydberg states~\cite{Nubbemeyer08,*eichmann2009,*Eichmann13,Landsman2013,Wolter14,*li14,*liu12,Diesen2016}.

In our experiment [Fig.\ \ref{fig:peda}], we use a reaction microscope to perform coincidence measurements of electrons and ions separated by the interaction of atoms with the laser and the weak dc fields~\cite{ullrich03,*doerner00}. The ionization signal of Rydberg states can be well distinguished of that from prompt laser induced strong-field ionization and retains a very high signal-to-noise ratio. Details of the experimental setup can be found in our previous publications~\cite{xie12prl,xie12_2}. Measurements were done with OTC laser fields formed by the superposition of a fundamental pulse with a center wavelength of 800 nm and its second harmonic with pulse durations (FWHM) of 46 fs and 48 fs, respectively. Temporal overlap of the two pulses was ensured by compensating their different group velocities with calcite plates and a pair of fused silica wedges. The electric field of the OTC pulses can be written as $\vec{E}(t,\Delta \varphi)=f_{x}(t)\cos(\omega t)\mathbf{\hat{e}_x}+f_{z}(t)\cos(2\omega t+\Delta \varphi)\mathbf{\hat{e}_z}$, with $\Delta \varphi$ the relative phase of the two colors and $f_{x,z}$ the pulse envelopes. The waveform of the OTC pulse can be precisely controlled on a sub-cycle time-scale via adjusting the position of one of the wedges. The peak laser intensity was about $6\times 10^{13}$ \wpcm\ (peak electric field on the order of $2\times 10^{8}$ V/cm) for each color. A weak homogeneous dc field of 1.5 V/cm was applied in the time-of-flight (TOF) spectrometer (along the polarization direction of the 400 nm pulse) to accelerate charged particles towards the detectors. This field also induces field ionization of high-lying Rydberg states populated during the strong field-atom interaction~\cite{Larimian2016}. A homogeneous magnetic field of 12 gauss is applied to ensure 4$\pi$ detection of electrons with velocities $v < 1.93$ a.u.\ ($E_\mathrm{kin} < 50.7$ eV). With our reaction microscope it is possible to observe electron-ion coincidences both from direct ionization events during the duration of the laser pulse and from delayed emission up to 20 $\mu$s after the pulse.

\begin{figure}[htbp]
\centering
\includegraphics[width=0.4\textwidth,angle=0]{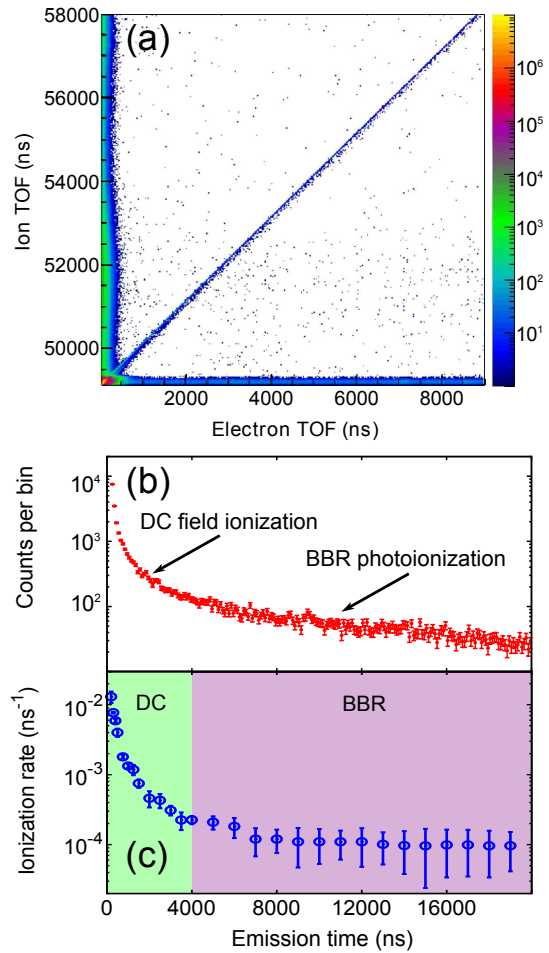}
\caption{(color online). (a) Photoelectron-photoion-coincidence (PEPICO) distribution for argon. The peak laser field strengths of both colors are $2\times10^8$ V/cm and the dc spectrometer field strength is 1.5 V/cm. (b) Signal of high-lying Rydberg states as a function of electron emission time [intensity distribution along diagonal in panel (a)]. (c) Ionization rate of Rydberg states derived from the data in panel (b).} \label{fig:pepico}
\end{figure}

The strong laser fields induce not only tunneling ionization of argon atoms but may also excite them to long-lived high-lying Rydberg states through electron recapture~\cite{Nubbemeyer08,*eichmann2009,*Eichmann13,Larimian2016,Diesen2016}. These high-lying Rydberg states can be ionized by a very weak dc field through over-the-barrier or tunneling ionization~\cite{Morishita2013} or through photoionization by photons absorbed from black-body radiation (BBR)~\cite{gallagher1979interactions}. A typical photo-electron photo-ion coincidence (PEPICO) distribution for argon interacting with an intense OTC field is shown in Fig.~\ref{fig:pepico}(a).
The delayed ionization signal from high-lying Rydberg atoms appears along the diagonal and can be easily separated from the prompt strong field ionization signal. From the correlated TOF signal between electrons and argon ions the emission time of the electrons from Rydberg atoms is extracted. Fig.~\ref{fig:pepico}(b) shows emission times of up to 20 $\mu$s. The ionization signal of high-lying Rydberg states contains two main contributions, one from field ionization by the weak dc extraction field applied along the spectrometer direction and the other one from photoionization induced by BBR at room temperature with a photon energy of about 0.025 eV~\cite{Spencer1982,*Beterov2009}. The (relatively) rapidly decaying ($\lesssim 1 \mu$s) component of the signal results from Rydberg states very close to the continuum threshold and well above the potential barrier field ionized by the weak dc field of $V_\mathrm{dc}=1.5$ V/cm [diabatic field ionization threshold $F_\mathrm{dc} = 1/(9 n_F^4)$ yielding $n_F\simeq 140$]. BBR photoionization from Rydberg states with $n>n_{BBR}\simeq 20$ mainly contributes to the slow decay with emission times longer than 4 $\mu$s. From the measured Rydberg signal, the corresponding ionization rates $\Gamma=-d\ln[I(\tau)]/d\tau$ can be derived [Fig.\ \ref{fig:pepico}(c)]. The rate decreases from about $\Gamma\approx 0.01$ ns$^{-1}$ at $\tau=200$ ns to about $2 \times 10^{-4}$ ns$^{-1}$ at $\tau=5$ $\mu$s. For emission times longer than 6 $\mu$s the ionization rate becomes nearly constant with a value of $\Gamma\approx 1 \times 10^{-4}$ ns$^{-1}$ which agrees well with the simulated photonionization rate by BBR \cite{Larimian2016}.

To analyze the formation of high-lying Rydberg atoms in the presence of OTC fields, we performed semiclassical electron ensemble simulations. Briefly, in the model the tunneling rate for strong-field ionization is derived from Landau's effective potential theory. The tunneled electrons are assumed to have a  Gaussian-like distribution over the transverse momentum perpendicular to the instantaneous laser field and zero longitudinal momentum along the instantaneous laser field. Each launched electron trajectory is weighted by the ADK ionization rate \cite{ADK}
and the initial lateral momentum distribution
\begin{equation}
%{W_1}({v_ \bot }^i) \propto [\sqrt {2{I_p}} /\left| {E({t_0})} \right|]\exp \left[ {\sqrt {2{I_p}} {{({v_ \bot }^i)}^2}/|E({t_0})|} \right]\, .
{W_1}({v_ \bot }^i) \propto \frac{\sqrt {2{I_p}}}{\left| {E(t_0)} \right|}\exp \left[-\frac{{\sqrt {2{I_p}} {{({v_ \bot }^i)}^2}}}{|E({t_0})|} \right]\, .
\end{equation}
${E}\left( t_0\right)$ is the laser electric field strength and $ I_p $ is the ionization potential. After tunneling and until the laser has concluded the classical Newtonian equations of motion ${\ddot{\vec r}} = - {\vec{r}}/{r^3} - {\vec{E}}(t)$ in the combined laser and Coulomb fields are solved numerically, where $r$ is the distance between the electron to the nucleus. For electrons with positive energy reaching the detector we calculate the asymptotic final momentum using Kepler's formula. Electrons with negative total energy are considered to be trapped in Rydberg states with large principal quantum numbers $n$. $7.2 \times {10^8}$ trajectories are simulated for Ar atoms at an intensity of $6\times 10^{13}$ \wpcm ~for each color. We use a trapezoidal envelope function with one optical cycle ramping up and down with four optical cycles in the plateau for the 800~nm laser pulse. The same envelope was used for the 400 nm laser pulse. For comparability with the experiment we consider only trajectories with final energies small enough to ensure $4\pi$ detection ($E_\mathrm{kin} < 50.7$ eV).

\begin{figure}[htbp]
\centering
\includegraphics[width=0.4\textwidth,angle=0]{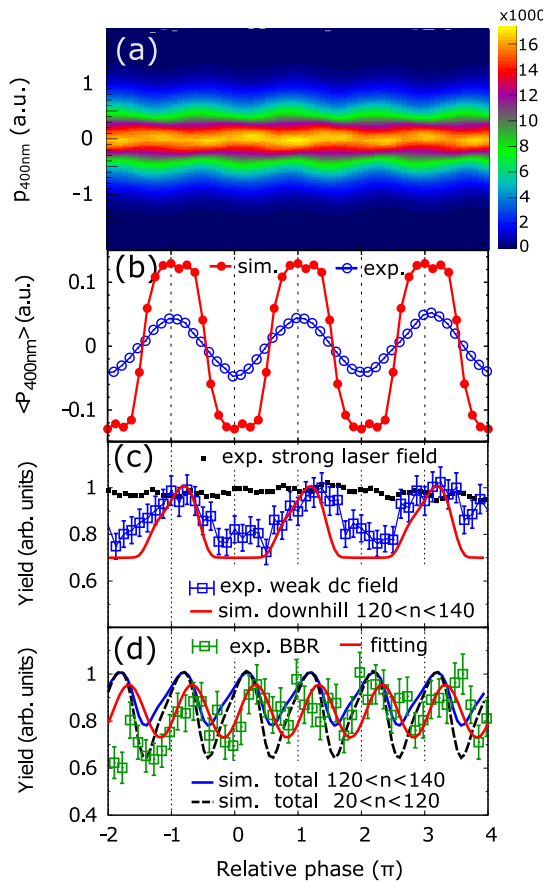}
\caption{(color online). (a) Measured momentum distribution of Ar$^+$ along the polarization direction of the 400 nm pulse as a function of the relative phase $\Delta\varphi$ between the two color fields. (b) The measured (blue circles) and simulated (red circles) average momentum of Ar$^+$ in the polarization direction of the 400 nm pulse as a function of the relative phase between the two color fields. (c) Normalized experimental strong field ionization yield (black squares) and dc-field ionization yield (blue squares with error bars) from high-lying Rydberg states as a function of the two-color phase. The red line is the projection of the dipole moment on in the ``downhill'' direction on the polarization axis of the 400 nm pulse. (d) Normalized BBR photoionization yield (green squares) with a fitting curve (red solid line) with a function of $0.838(\pm1.4\%)+0.117(\pm25\%)\cos[2\pi\Delta\varphi-0.64\pi(\pm3.4\%)]$ and simulated population of Rydberg states with $20<n<120$ (black dashed line) and $120<n<140$ (blue solid line) as a function of the relative phase. In all panels the absolute value of the two-color phase has been determined by matching the oscillations of the average momentum in panel (b).} \label{fig:delay_scanning}
\end{figure}
The final momentum of the ions and electrons from the strong field ionization is determined by the vector potential of the external laser field at the ionization time. In the case of an OTC laser field the cycle waveform changes periodically with the relative phase between the two laser components~\cite{zhang14prl} leading to a periodic modulation of the momentum distribution of the argon ions with the relative phase. In Fig.~\ref{fig:delay_scanning}(a) the measured momentum distribution of Ar$^+$ along the polarization axis of the 400 nm pulse is shown as a function of the relative phase. The clear periodic dependence of the momentum distribution on the relative phase indicates that a precise control of the cycle shape of the OTC field with the relative phase is achieved in the experiments. The average momentum of Ar$^+$ ions along the polarization axis of the 400 nm pulse is plotted together with the simulated values in Fig.~\ref{fig:delay_scanning}(b). Both the measured and the simulated data oscillate periodically with the relative phase. The experimental curve, however, features a smaller oscillation amplitude and a different shape which could result from averaging effect due to the experimental pulse intensity profile near the laser focus in the interaction region which is not accounted for in our simulation.

It is now instructive to analyze delayed dc field ionization and BBR photoionization separately. First, we select the dc field ionization yield of high-lying Rydberg states by integrating the signal over emission times in the emission-time interval between 100 ns and 4 $\mu$s after the laser pulse. This yield [blue squares with error bars in Fig.~\ref{fig:delay_scanning}(c)] exhibits a clear $2\pi$-periodicity which is well reproduced by our simulation (red solid line). It is important to note that this $2\pi$-periodicity is neither related to the strong-field ionization yield (black dots) which is almost constant for all relative phases nor to the electron recapture rate [blue line in Fig.\ \ref{fig:delay_scanning}(d)] which oscillates with $\pi$-periodicity. A  $\pi$-periodicity is also found for the photoionization yield by BBR for electron emission times longer than 4 $\mu$s [green squares with error bars in Fig.~\ref{fig:delay_scanning}(d)]. The origin of the observed different periodicities lies in the localization of the Rydberg wave packet as revealed by our simulations.

Starting point is the inversion symmetry of the atom and the $\pi$-periodicity of the laser intensity $|\vec E(\Delta\varphi)|^2 = |\vec E(\Delta\varphi+\pi)|^2$. Therefore, all processes that do not depend on the directionality of the field, including the formation of high lying Rydberg states, will feature the same periodicity as the intensity of the laser field. This is indeed also observed in our simulation when counting the number of electrons with negative final energy [the blue solid line and the black dashed line in Fig.\ \ref{fig:delay_scanning}(d)]. As also black-body radiation is isotropic, the yield of post-pulse photoionized Rydberg atoms is directly proportional to the number of available highly excited atoms and, consequently, exhibits the same $\pi$-periodicity [Fig.\ \ref{fig:delay_scanning} (d)].

\begin{figure}[htbp]
\centering
\includegraphics[width=0.4\textwidth,angle=0]{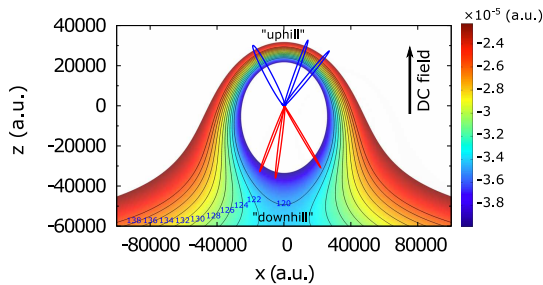}
\caption{(color online) Atomic potential in a weak dc field with field strength $F_\mathrm{dc}=1.5$ V/cm. The contour lines denote the energy levels of hydrogenic Rydberg states with their principle quantum numbers indicated. The red and blue ellipses represent orbitals of red- and blue-shifted Rydberg-Stark states, respectively. The arrow indicates the direction of the dc field.} \label{fig:pot}
\end{figure}
The situation is different for the post-pulse dc field ionization where the weak extraction field present in the interaction region breaks the inversion symmetry of the system. As shown in Fig.~\ref{fig:pot}, red-shifted Stark states with elongated orbitals pointing in the ``downhill'' direction, i.e., towards the potential barrier formed by the Coulomb and the static fields have the highest probability to overcome the barrier even though blue-shifted states (``uphill'') states are energetically higher~\cite{Larimian2016}. We thus use the average dipole moment of the electrons or, equivalently, the spatial localization of the charge cloud, as an indicator  for the probability of the ensemble to escape due to the post-pulse interaction with the weak dc field. To this end we analyze the dipole moment of all states with (hydrogenic) primary quantum number $120<n<140$, i.e., with energies close to or above the potential barrier. Remarkably, we find the average dipole moment to be directly correlated with the relative phase between the two color components. Consequently, although the dipole moment is aligned along the dc (or 400 nm laser) field axis with $\pi$-periodicity, the orbit is elongated in downhill direction only with $2\pi$-periodicity resulting in an increased ionization yield due to the presence of the weak dc field. The red line in Fig.\ \ref{fig:delay_scanning}(c) shows the projection of the dipole moment in the ``downhill'' direction on the polarization axis of the 400 nm pulse as a function of the two-color phase. Interestingly, the oscillation is not symmetric for uphill and downhill directions but, instead, shows a ``dip'' in the uphill direction due to the influence of the static field on the orbit of the recaptured electron. The flat uphill part of the oscillation almost perfectly reproduces the measured post-pulse dc field ionization yield [blue open squares in Fig.\ \ref{fig:delay_scanning} (c)]. The agreement between simulated and measured emission yields points to the high degree of spatial localization of states in one hemisphere which can be populated in laser-atom interactions by controlling the waveform of the exciting laser pulse.

In conclusion, we have presented a joint experimental and theoretical study on highly excited Rydberg states created during the interaction of argon atoms with waveform controlled OTC laser fields. The ionization yields due to weak dc field ionization and BBR photoionization could be measured separately as a function of the relative phase between the two colors. We found different oscillation periods of 2$\pi$ and $\pi$, respectively. The measurements and analysis of trajectories calculated in a semiclassical simulation suggest that Rydberg electrons recaptured by the ion after conclusion of the pulse can be preferentially steered and localized into a hemisphere of the atom by the shape of the laser field waveform. We have demonstrated this steering in our experiment by manipulating the relative phase between the two colors. For specific two-color phases, Rydberg wave packets with energies above the potential barrier are predominantly localized on the downhill side of the atom. The ionization yield of such electrons induced by a weak dc field is significantly increased and shows a modulation with a period of $2\pi$. The present study provides an effective way to control the population and the localization of high-lying Rydberg wave packets and may find potential applications in the manipulation of interacting Rydberg ensembles, Rydberg molecules, and chemistry.

\acknowledgments
This work was financed by the Austrian Science Fund (FWF) under P25615-N27, P28475-N27, P21463-N22 and P23359-N16, special research programmes SFB-041 ViCoM, SFB-049 NextLite, and doctoral college W1243, by the National Natural Science Foundation of China~(NNSFC) under Grant No.~11574010, and by  the National Program on Key Basic Research Project~(973 Program) under Grant No.~2013CB922402.

\end{document}